\documentclass{moriond}

\def\be{\begin{equation}}
\def\ee{\end{equation}}
\def\bea{\begin{eqnarray}}
\def\eea{\end{eqnarray}}

\newcommand{\Photo}{}

\begin{document}
\vspace*{4cm}
\title{SO(3) FAMILY SYMMETRY AND AXIONS}

\author{MARIO REIG}

\address{AHEP Group, Institut de F\'{i}sica Corpuscular --
	CSIC/Universitat de Val\`{e}ncia, Parc Cient\'ific de Paterna.\\
	C/ Catedr\'atico Jos\'e Beltr\'an, 2 E-46980 Paterna (Valencia) - SPAIN}

\maketitle\abstracts{Motivated by the idea of Comprehensive Unification, we consider a gauged $SO(3)$ flavor extension of the Standard Model, including right-handed neutrinos and a Peccei-Quinn symmetry. The model accommodates the observed fermion masses and mixings and yields a characteristic, successful relation among them. The Peccei-Quinn symmetry is an essential ingredient.}

\section{$SO(3)$ as a gauge family symmetry: the threefold way}
Since the late '70s and beginning of the '80s many proposals have been made to explain family replication and the pattern of fermion masses and mixings. To this end, it seems appropriate to consider symmetry groups containing triplet representations. Many possibilities emerge. For example, one can use discrete symmetries like $\Delta (27)$ \cite{deMedeirosVarzielas:2017sdv}, $A_4$ \cite{Morisi:2011pt} or $T_7$ \cite{Bonilla:2014xla} as flavor symmetries since all of them contain triplet representations. However, only two options appear if one considers continuous symmetries: $SU(3)$ and $SO(3)$. Therefore, the requirement of a gauged theory of flavor reduces our possible choices considerably.

$SU(3)$ is an appealing possibility that has been studied in the past (see for example  \cite{Berezhiani:1989fp}). This family symmetry is particularly interesting because one can use Higgses in sextet, $\mathbf{6}$, representation. The $\mathbf{6}$ is an interesting representation to explain why the third family is much heavier than the second and first generations. A fermion mass term, in this case, must come from the vacuum expectation value (VEV) of a Higgs with $SU(3)$ charge
\begin{equation}
M_f\sim y_f\bar{f}_L\langle H \rangle f_R\,.
\end{equation}
This possibility, however, requires a chiral assignment of flavor charge to fermions, $f_L\sim \mathbf{3}$ and $f_R\sim \mathbf{3^*}$, and is plagued by anomalies unless extra, exotic fermions are introduced to cancel them. Therefore one loses minimality to generate the hierarchy among generations \footnote{Note that this requirement of anomaly cancellation is absent if the $SU(3)_F$ is a global symmetry. However, quantum gravity might require that all symmetries should be gauged \cite{Krauss:1988zc}.}. 
We will not consider this possibility further. 

In this work we consider $SO(3)$ as a gauge family  symmetry \cite{Reig:2018ocz}. The $SO(3)$ group is theoretically interesting because it is more easily compatible with the ideas of Grand Unified Theories (GUT). For example, in the usual $SU(5)$ and $SO(10)$ theories one embeds the Standard Model particle content in the chiral, anomaly free sets of representations:
	$3\times (\mathbf{\bar{5}}+\mathbf{10})$ for $SU(5)$ and 
	$3\times\mathbf{16}$ for $SO(10)$. 
	As we have said, assigning these representations as $SU(3)$ triplets generally leads to anomalies. For example, in the $SO(10)\times SU(3)$ theory the standard $(\mathbf{16},\mathbf{3})$ combination has an $[SU(3)_F]^3$ anomaly. This is not the case for $SO(3)$ because it is automatically anomaly free. 
	This property was used in \cite{Reig:2017nrz} where we revived the idea of Comprehensive Unification, merging gauge and family symmetry, that was initially proposed in the '80s \cite{Wilczek:1981iz}. More specifically, the breaking scheme $SO(18) \rightarrow SO(10) \times SO(5) \times SO(3)$
	\cite{GellMann:1980vs} allows for the standard $SO(10)$ gauge unification together with a hypercolor $SO(5)$, which confines the 5 extra families (leaving 3), and an $SO(3)$ family symmetry group.  This motivates consideration of $SO(3)$ as a family unification group.  
	$SO(3)$ as a gauge family symmetry was first proposed in \cite{Wilczek:1978xi}. In their pioneer work, Wilczek and Zee proposed that in the same way the $SU(2)$ group relates up and down-type fermions a new interaction relating families in the \textit{horizontal} direction could explain family replication. In addition the authors also propose a particular symmetry breaking pattern as the origin of the fermion mass and mixing hierarchies. 
	
	The use of $SO(3)$ family symmetry was a successful, predictive scenario. First of all, in that framework quark mixing angles can be written in terms of quark masses:
	\begin{equation}
	\label{CKM_Wilczek}
	V_{CKM}\approx\left(\begin{array}{ccc}
	1 & -\sqrt{\frac{m_d}{m_s}}+\sqrt{\frac{m_u}{m_c}} & \sqrt{\frac{m_um_c}{m_t^2}}-\sqrt{\frac{m_dm_s}{m_b^2}} \\
	\sqrt{\frac{m_d}{m_s}}-\sqrt{\frac{m_u}{m_c}} & 1 & 0 \\
	-\sqrt{\frac{m_um_c}{m_t^2}}+\sqrt{\frac{m_dm_s}{m_b^2}}  & 0 & 1
	\end{array}\right)\,.
	\end{equation} 
	In addition they predicted an interesting relation between quark and lepton masses
	\begin{equation}
	\frac{m_em_\mu}{m_\tau^2}=\frac{m_dm_s}{m_b^2}=\frac{m_um_c}{m_t^2}\,.
	\end{equation}
	With this formula, the threefold way was able to predict a very light top quark:
	\begin{equation}
	m_{t}^{predicted}\approx 15\,\,GeV.
	\end{equation}
	It is important to notice that this idea was born 16 years before the top quark was discovered. In 1995 the top quark was discovered with a mass around 173 GeV, ruling out the original threefold way. Notice also that due to the predicted CKM matrix (see Eq. \ref{CKM_Wilczek}) the $b$ quark, whose properties were not very well known by that time, was expected to decay mainly to up quarks in the original $SO(3)$ scenario.

\section{$SO(3)\times U(1)_{PQ}$, the threefold way revamped}
In this work \cite{Reig:2018ocz} we revisit the $SO(3)$ family symmetry scenario and study which are the requirements to make it phenomenologically viable.
We find that the implementation of the PQ mechanism \cite{Peccei:1977hh,Wilczek:1977pj,Weinberg:1977ma} \textit{\`a la DFSZ} \cite{Dine:1981rt} allows to avoid the wrong top quark mass prediction and solve the strong CP problem, simultaneously. A crucial point is that in the absence of heavy vector-like quarks the QCD anomaly condition, needed to implement the PQ mechanism, can only be achieved by introducing a duplicated Higgs sector (see table \ref{tab:content}). We assume the following pattern for their VEV
\begin{equation}\label{ssb_3s}
\langle\Phi^{u,d}\rangle =
\left (\begin{array}{ccc}
0&    & \\
& -k^{u,d} & \epsilon_1^{u,d} \\
&  \epsilon_1^{u,d}   &\hspace{1mm}\,\,k^{u,d} \\
\end{array}\right)\,,\,\,\,\,\,
\langle \Psi^{u,d}\rangle =
\left (\begin{array}{ccc}
v^{u,d} \\
0 \\
\epsilon_2^{u,d} \\
\end{array}\right)~.
\end{equation} 
The $\epsilon_{1,2}$ correspond to small perturbations around the minimum of the scalar potential. These scalars will couple to up-type and down-type fermions selectively. 
\begin{equation}\label{eq:yuklag}
\mathcal{L}=\bar{q}_L(y_1\Psi^u+y_2\Phi^u)u_R+\bar{q}_L(y_3\Psi^d+y_4\Phi^d)d_R+\bar{l}_L(y_5\Psi^d+y_6\Phi^d)e_R+h.c.
\end{equation}
After electroweak breaking,
Eq.~(\ref{eq:yuklag}) leads to the quark mass matrices
\begin{equation}
\label{mass.u}
M^u=\left(\begin{array}{ccc}
0 & y_1\epsilon^u_2 & 0 \\
-y_1\epsilon^u_2 & -y_2k^u & y_1v^u + y_2\epsilon_1^u \\
0 & -y_1v^u + y_2\epsilon_1^u & y_2k^u
\end{array}\right)\,,\,\,\,\,
M^d=\left(\begin{array}{ccc}
0 & y_3\epsilon^d_2 & 0 \\
-y_3\epsilon^d_2 & -y_4k^d & y_3v^d + y_4\epsilon_1^d \\
0 & -y_3v^d+ y_4\epsilon_1^d & y_4k^d
\end{array}\right),
\end{equation}
while for the charged leptons we have
\begin{equation}
\label{mass.e}
M^e=\left(\begin{array}{ccc}
0 & y_5\epsilon^d_2 & 0 \\
-y_5\epsilon^d_2 & -y_6k^d & y_5v^d+ y_6\epsilon_1^d \\
0 & -y_5v^d+ y_6\epsilon_1^d & y_6k^d
\end{array}\right)~,
\end{equation}
where we took into account the VEV alignment patterns of the $SO(3)$ triplet and quintuplet scalars, respectively (see \cite{Reig:2018ocz} for more details). From the mass matrices above we get, in first approximation neglecting the perturbations $\epsilon_{1,2}$, the fermion masses 
\begin{equation}\label{masses}
m_{u,d, e}=0\,,\,\,\,\,\,\,\,\,
m_{c,s,\mu}=|y_{2,4,6}k^{u,d}-y_{1,3,5}v^{u,d}|\,,\,\,\,\,\,\,\,\,
m_{t,b,\tau}=|y_{2,4,6}k^{u,d}+y_{1,3,5}v^{u,d}|\,.
\end{equation}
Once the perturbations $\epsilon_{1,2}^{u,d}$ are taken into account a small mass is generated for the first family
\begin{equation}
m_{1st}\sim \epsilon_2^2/m_{2nd}\,.
\end{equation}
One can notice that this resembles to a seesaw mass relation and explains the smallness of the first family. In addition, turning on the perturbations, $\epsilon_i$, allows us to get the conceptual relation between quark and lepton masses (also present in a $A_4$ supersymmetric scenario \cite{Morisi:2011pt})
\begin{equation}
\label{eq:gold}
\frac{m_\tau}{\sqrt{m_em_\mu}}\approx\frac{m_b}{\sqrt{m_dm_s}}~.
\end{equation}
This successful formula nicely relates down-type quark and
charged lepton masses.  On the other hand, the doubled Higgs structure forced by PQ symmetry allows us to avoid  the unwanted top quark
mass prediction $\frac{m_\tau}{\sqrt{m_em_\mu}}\approx\frac{m_t}{\sqrt{m_um_c}}$ present in~\cite{Wilczek:1978xi}. In addition to this relation, quark mixing angles can be written in terms of quark masses with a reasonable accuracy
\begin{equation}
\theta_C\approx \sqrt{\frac{m_d}{m_s}}- \sqrt{\frac{m_u}{m_c}}\,,\,\,|V_{ub}|\approx \frac{\sqrt{m_dm_s}}{m_b}- \frac{\sqrt{m_um_c}}{m_t}\,,\,\,|V_{cb}|\approx \frac{\epsilon_1^u}{2k^u}-\frac{\epsilon^d_1}{2k^d}\,.\,\
\end{equation}
Notice that unlike $\theta_C$ and $|V_{ub}|$, the $|V_{cb}|$ matrix element is not given in terms of quark masses but it is predicted to be small. This solves the issue of the original threefold way \cite{Wilczek:1978xi}, that predicted the $b$ quark decaying mainly to up quarks through the weak charged current. 

	\begin{table}[t]
		\caption{Particle content and transformation properties under the SM
			and flavor $SO(3)$ gauge groups. The VEVs of SM singlets $\sigma$ and
			$\rho$ break $U(1)_{PQ}$ and lepton number, generating Majorana
			neutrino masses.}
		\label{tab:content}
		\begin{center}
			\begin{tabular}{|c|c|c|c|c|c|c|c|c|c|c|c|c|}
				\hline
				& $q_L$ &$u_R$&$d_R$ &$l_L$ &$e_R$ &$\nu_R$ & $\Phi^u$ & $\Phi^d$ & $\Psi^u$ & $\Psi^d$ & $\sigma$ & $\rho$\\
				\hline
				$\mathrm{SU(3)_c}$ & \textbf{3} &\textbf{3} &\textbf{3} &\textbf{1} &\textbf{1} &\textbf{1}  & \textbf{1} & \textbf{1} & \textbf{1} &\textbf{1} & \textbf{1} & \textbf{1} \\
				\hline
				$\mathrm{SU(2)_L}$ & \textbf{2} & \textbf{1} & \textbf{1}  & \textbf{2} & \textbf{1} & \textbf{1} & \textbf{2} & \textbf{2} & \textbf{2} & \textbf{2}  &\textbf{1} &\textbf{1}\\
				\hline
				$\mathrm{U(1)_{Y}}$ & $\frac{1}{6}$ & $\frac{2}{3}$  & -$\frac{1}{3}$  & -$\frac{1}{2}$ & $-1$ & $0$ & -$\frac{1}{2}$ & $\frac{1}{2}$ & -$\frac{1}{2}$& $\frac{1}{2}$  &$0$ & $0$\\ 
				\hline
				$\mathrm{SO(3)_F}$ & \textbf{3} &\textbf{3} &\textbf{3} &\textbf{3} &\textbf{3} & \textbf{3} & \textbf{5} & \textbf{5} & \textbf{3} & \textbf{3}  &$\mathbf{5}$&$\textbf{1}$\\
				\hline
				$\mathrm{U(1)_{PQ}}$ & 1 & -1 & -1 & {1} & {-1} & {-1} & 2 & 2 & 2 & 2 & {2} & {2}\\
				\hline
			\end{tabular}
		\end{center}
	\end{table}

Another interesting feature of the framework is that since lepton number, PQ and $SO(3)$ family symmetries are spontaneously broken at the same scale some connections between flavor, neutrino mass and the axion mass scales appear. In particular, the axion and neutrino mass scales get connected through the conceptual relation
\begin{equation}
m_a\sim \left(\Lambda_{QCD}m_\pi/v_{EW}^2\right)m_\nu\,.
\end{equation}
The connection between family and PQ symmetry breaking scales allows one to constrain the axion decay constant. This is because the $SO(3)$ gauge bosons can mediate $\Delta F=2$ processes like $K^0-\bar{K}^0$ mixing, which constrain the gauge boson mass to be
\begin{equation}
\frac{g^2}{M_F^2}\leq \frac{1}{[10^4\, TeV]^2}\,.
\end{equation} 
Since $SO(3)$ bosons obtain their mass from the scalar breaking PQ symmetry, a constrain to the axion decay constant appears: $f_a\geq 10^7$ GeV. This lower bound is, however, much weaker than astrophysical constraints \cite{Raffelt:2006cw}.

\section{Conclusions}
Taking Comprehensive Unification using spinors as a guideline we have studied the consequences of extending the Standard Model with an $SO(3)$ family symmetry. We find that the implementation of the PQ symmetry is a crucial ingredient to achieve a phenomenologically viable theory which also offers interesting predictions in the flavor sector. In addition, the predicted QCD axion constitutes, as usual, a successful dark matter candidate. 

\section*{Acknowledgments}
I would like to thank the organizers of the 54th Rencontres de Moriond for arranging a very interesting conference. I also want to thank Jose W.F. Valle and Frank Wilczek for a stimulating collaboration. This work is funded by the Spanish grants SEV-2014-0398 and FPA2017-85216-P (AEI/FEDER, UE) and PROMETEO/2018/165 (Generalitat Valenciana) and by FPU grant FPU16/01907.


\section*{References}

\begin{thebibliography}{99}

\bibitem{Reig:2018ocz}
M.~Reig, J.~W.~F.~Valle and F.~Wilczek,
Phys.\ Rev.\ D {\bf 98} (2018) no.9,  095008
doi:10.1103/PhysRevD.98.095008
[arXiv:1805.08048 [hep-ph]].

\bibitem{deMedeirosVarzielas:2017sdv}
I.~de Medeiros Varzielas, G.~G.~Ross and J.~Talbert,
JHEP {\bf 1803} (2018) 007
doi:10.1007/JHEP03(2018)007
[arXiv:1710.01741 [hep-ph]].

\bibitem{Morisi:2011pt}
S.~Morisi, E.~Peinado, Y.~Shimizu and J.~W.~F.~Valle,
Phys.\ Rev.\ D {\bf 84} (2011) 036003
doi:10.1103/PhysRevD.84.036003
[arXiv:1104.1633 [hep-ph]].

\bibitem{Bonilla:2014xla}
C.~Bonilla, S.~Morisi, E.~Peinado and J.~W.~F.~Valle,
Phys.\ Lett.\ B {\bf 742} (2015) 99
doi:10.1016/j.physletb.2015.01.017
[arXiv:1411.4883 [hep-ph]].

\bibitem{Krauss:1988zc}
L.~M.~Krauss and F.~Wilczek,
Phys.\ Rev.\ Lett.\  {\bf 62} (1989) 1221.
doi:10.1103/PhysRevLett.62.1221

\bibitem{Berezhiani:1989fp}
Z.~G.~Berezhiani and M.~Y.~Khlopov,
Z.\ Phys.\ C {\bf 49} (1991) 73.
doi:10.1007/BF01570798

\bibitem{Reig:2017nrz}
M.~Reig, J.~W.~F.~Valle, C.~A.~Vaquera-Araujo and F.~Wilczek,
Phys.\ Lett.\ B {\bf 774} (2017) 667
doi:10.1016/j.physletb.2017.10.038
[arXiv:1706.03116 [hep-ph]].

\bibitem{Wilczek:1981iz}
F.~Wilczek and A.~Zee,
Phys.\ Rev.\ D {\bf 25} (1982) 553.
doi:10.1103/PhysRevD.25.553

\bibitem{GellMann:1980vs}
M.~Gell-Mann, P.~Ramond and R.~Slansky,
Conf.\ Proc.\ C {\bf 790927} (1979) 315
[arXiv:1306.4669 [hep-th]].

\bibitem{Wilczek:1978xi}
F.~Wilczek and A.~Zee,
Phys.\ Rev.\ Lett.\  {\bf 42} (1979) 421.
doi:10.1103/PhysRevLett.42.421

\bibitem{Peccei:1977hh}
R.~D.~Peccei and H.~R.~Quinn,
Phys.\ Rev.\ Lett.\  {\bf 38} (1977) 1440.
doi:10.1103/PhysRevLett.38.1440

\bibitem{Wilczek:1977pj}
F.~Wilczek,
Phys.\ Rev.\ Lett.\  {\bf 40} (1978) 279.
doi:10.1103/PhysRevLett.40.279

\bibitem{Weinberg:1977ma}
S.~Weinberg,
Phys.\ Rev.\ Lett.\  {\bf 40} (1978) 223.
doi:10.1103/PhysRevLett.40.223

\bibitem{Dine:1981rt}
M.~Dine, W.~Fischler and M.~Srednicki,
Phys.\ Lett.\  {\bf 104B} (1981) 199.
doi:10.1016/0370-2693(81)90590-6

\bibitem{Raffelt:2006cw}
G.~G.~Raffelt,
Lect.\ Notes Phys.\  {\bf 741} (2008) 51
doi:10.1007/978-3-540-73518-2-3
[hep-ph/0611350].

\end{thebibliography}
\end{document}